\documentclass[conference]{IEEEtran}
\IEEEoverridecommandlockouts
\usepackage{cite}
\usepackage{amsmath,amssymb,amsfonts}
\usepackage{algorithmic}
\usepackage{graphicx}
\usepackage{textcomp}
\usepackage{xcolor}
\usepackage[hyphens]{url}
\def\BibTeX{{\rm B\kern-.05em{\sc i\kern-.025em b}\kern-.08em
    T\kern-.1667em\lower.7ex\hbox{E}\kern-.125emX}}

\makeatletter
\newcommand{\figcaption}[1]{\def\@captype{figure}\caption{#1}}
\newcommand{\tblcaption}[1]{\def\@captype{table}\caption{#1}}
\makeatother
\begin{document}

\title{Data Backup System with No Impact on Business Processing Utilizing Storage and Container Technologies
}

\author{\IEEEauthorblockN{Satoru Watanabe}
	\IEEEauthorblockA{\textit{Hitachi, Ltd.} 
		\textit{Research \& Development Division} \\
		Kokubunji-shi, Tokyo, Japan \\
		satoru.watanabe.aw@hitachi.com}
	}

\maketitle

\begin{abstract}
  Data backup is a core technology for improving system resilience to system failures.
Data backup in enterprise systems is required to minimize the impacts on business processing, which can be categorized into two factors: system slowdown and downtime.
To eliminate system slowdown, asynchronous data copy (ADC) technology is prevalent, which copies data asynchronously with original data updates. 
However, the ADC can collapse backup data when applied to enterprise systems with multiple resources. 
Then, the demonstration system employed consistency group technology, which makes the order of data updates the same between the original and backup data.
In addition, we developed a container platform operator to unravel the complicated correspondence between storage volumes and applications. 
The operator automates the configuration of the ADC with the setting of consistency groups.
We integrated the storage and container technologies into the demonstration system, which can eliminate both system slowdown and downtime. 
\end{abstract}


\section{Introduction}\label{intro}
Data backup is a core technology for improving system resilience to system failures.
System failure can be caused by various types of incidents, such as hardware malfunctions, software bugs, misoperations, cyber-attacks, and natural disasters.
These incidents can damage systems, and data backup protects the most important assets in the system or data. 

Data backup in enterprise systems is required to minimize the impact on business processing.
The impact on business processing can be categorized into two factors: system slowdown and downtime.
System slowdown impairs customer satisfaction \cite{gelderman1998relation}, and system downtime can cause business losses, reaching millions of US dollars per hour \cite{wang2020enterprise}.

To eliminate system slowdown, asynchronous data copy (ADC) is prevalent, which copies data asynchronously with the data updates at the main site \cite{pandey2014survey}.
The ADC can separate the processes of data updates at main and backup sites.
It is indispensable for removing the negative impact on system performance.

Elimination of system downtime is challenging for enterprise systems because of their complexity.
In an e-commerce system \cite{saona2021technologies}, business processing is executed using multiple resources such as inventory and payment databases.
When the ADC is applied to this system, the backup data can collapse, meaning that some transaction data are included in the inventory backup data but not in the payment backup data, and vice versa.
Thus, collapsed data are not adequate for recovering enterprise systems due to its disturbance in business process.

For preventing data collapse, the order of data update is essential.
Enterprise systems are designed to recover the system in case resources fail during business processing \cite{bernstein1990transaction}.
The recovery relies on the storage system, which protects data in the order of ``acks'' (acknowledgements to data update request) from storage to server.
For example, a storage system sent two ``acks'' one after another.
If the first and second ``acks'' were invalid and valid, respectively, recovery cannot work.
The recovery can work under the protection of the order of data updates.

Furthermore, disaster recovery (DR) systems with the ADC are designed to recover the backup site even when some ``acks'' are invalid \cite{patterson2002snapmirror, choy2000disaster}.
Owing to network delays, data loss at the backup site are inevitable.
However, DR systems recover the backup site under the condition of data consistency, where data consistency means that the order of data update is the same between the main and backup data \cite{brailey2009protecting}.
The storage system provides consistency group technology, which ensures data consistency between the main and backup data \cite{mikkelsen2005ensuring}.
This technology was employed in the demonstration system to eliminate system downtime.

In addition, we developed a container platform operator to determin the correspondence between storage volumes and applications.
Storage systems can include hundreds of volumes, which can be used in hundreds of applications.
Thus, the correspondence among data volumes and applications is highly complicated.
The operator automates the configuration of the ADC with the setting of consistency groups.
We integrated storage and container technologies into the demonstration system, which can eliminate both system slowdown and downtime. 

\begin{figure*}
	\centering
	\includegraphics[scale=1.0]{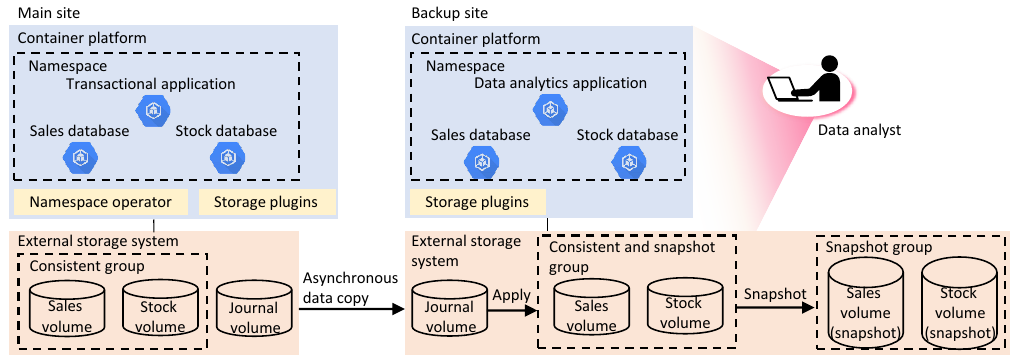}
	\caption{Use case and technologies overview in the demonstration system.}
	\label{fig1}
\end{figure*}

\section{Use Case}
Fig. \ref{fig1} illustrates the demonstration system consisting of main and backup sites.
Both have a container platform and an external storage system. 

Namespaces in the container platform partition the application environment for preventing them from affecting each other \cite{chang2017kubernetes}.
The demonstration assumes the business process works on a namespace.
The namespace includes the transactional application and two databases (sales and stock databases) for the business process.

Users operate the system using the consoles provided by the container platforms.
When users provide instructions for starting backup on the main site, the namespace operator identifies the data volumes related to the business process.
And the operator configures the ADC using the storage plugins.
The automation of the operator removes the laborious tasks to identify the related data volumes and to configure the ADC.
For utilizing backup data, the demonstration system employs snapshot technology, which duplicated volumes in the external storage system \cite{bertrand2004examining}.
The operations for configuring the snapshot are completed on the container platform in the backup site.

The container storage interface (CSI\footnote{https://github.com/container-storage-interface}) for snapshot group was introduced as an alpha feature\footnote{\url{https://kubernetes.io/blog/2023/05/08/kubernetes-1-27-volume-group-snapshot-alpha/}}.
The CSI standardizes the operations of external storage systems, which vary depending on the vendors of the storage system.
It is difficult for system administrators to be experienced with the various kinds of storage systems. 
The CSI removes the learning overhead of storage systems and enables utilizing the storage technologies without prior knowledge about the storage system.
Because it remains an alpha feature, the storage plugin for snapshot group has not yet supported the CSI.
Thus, for configuring snapshot groups, users need to operate the external storage system directly.
The storage operation will be removed by the technical advancements in the CSI and the storage plugin in the future. 

Owing to the consistent and snapshot group technologies, the data in the main and backup sites remain consistent.
The data consistency enables to recover the system in the backup site.

\section{Technologies}\label{technology}
This section explains the technologies employed in the demonstration system.
All the technologies, except for the namespace operator, have been commercialized. 

\subsection{Storage technologies}

\subsubsection{Asynchronous data copy}
The external storage system provides the ADC function using two journal volumes in the main and backup sites (Fig.\ \ref{fig1}).
When data are updated on the main site, the update logs are stored in the journal volume (main site), which are transferred to the journal volume (backup site) asynchronously.
Volumes in the backup site are updated according to the logs.
All the data in the main site are copied to the backup site at the ADC initialization.
Thus, the data volumes in the backup site are consistent, meaning that data in the main and backup sites are updated in the same order.

The external storage system also provides a consistency group function, which shares the journal volume with multiple volumes.
Consequently, the groups of volumes in the main and backup sites are retained in consistent, essential for recovering the backup site, as described in Section \ref{intro}.

\subsubsection{Snapshot}
The demonstration system employs snapshot technology in the backup site for utilizing the backup data.
Snapshot volumes store the original data when the volumes are updated as shown in Fig.\ \ref{fig1}.
Using the snapshot technology, the storage system can provide the updated and original volumes.
The demonstration system utilizes the updated and original volumes for data backup and analytics, respectively.

The external storage system also provides snapshot group technology, creating snapshots for multiple volumes simultaneously.
The snapshot group technology enables the demonstration system to retain the snapshot volumes in consistent with the volumes on the main site.

\subsection{Container technologies}

\subsubsection{Namespace operator}
Namespace operator (NSO) configures the ADC according to user operations on the console in the main site.
When users put a tag to the target namespace, the NSO extracts all the volumes in the namespace and creates custom resources for configuring the ADC and consistent group.
The custom resources control the storage plugins described in the next section.
We developed the NSO using the operator SDK\footnote{\url{https://docs.openshift.com/container-platform/4.13/operators/operator_sdk/osdk-about.html}} provided by RedHat.

\subsubsection{Storage plugin}\label{plugin}
For configuring the ADC, the demonstration system employs Replication Plug-in for Containers\footnote{\url{https://knowledge.hitachivantara.com/Documents/Adapters_and_Drivers/Storage_Adapters_and_Drivers/Containers/Replication_Plug-in_for_Containers/1.1.0/Replication_Plug-in_for_Containers_Configuration_Guide}} and Storage Plug-in for Containers\footnote{\url{https://knowledge.hitachivantara.com/Documents/Adapters_and_Drivers/Storage_Adapters_and_Drivers/Containers/Storage_Plug-in_for_Containers}}.
These plug-ins operate the storage systems on behalf of users.
By utilizing the plug-ins provided by storage vendors, users can operate storage systems through the common interface of the container platform.
Thus, these plug-ins remove the learning overhead of storage operations for utilizing storage technologies.

\section{Demonstration}

\subsection{Overview}
The demonstration system\footnote{The demonstration video is available at \url{https://photos.app.goo.gl/Qq8LdxHLeyuw285t6}.} comprises two Hitachi Virtual Storage Platform G370s, two Openshift platforms version 4.13, and four Oracle databases version 23c.
The main and backup sites contain one G370, one Openshift, and two Oracle databases each.
The two G370s are connected to a network for copying the data.
The system is operated using the web consoles provided by the Openshift platforms.

\begin{figure}
	\centering
	\includegraphics[width=\linewidth]{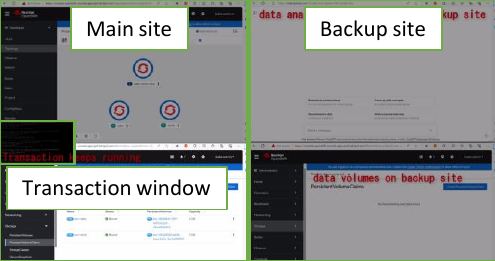}
	\caption{Screen structure of the demonstration.}
	\label{fig2}
\end{figure}

Fig.\ \ref{fig2} illustrates the screen structure of the demonstration.
The screen is divided vertically, wherein the left and right halves show the main and backup sites, respectively.
A transaction window is used in the left half for illustrating the continual execution of transactions.

The demonstration consists of three steps: backup configuration, snapshot development, and data analytics.
In the backup configuration step, the demonstration configures the ADC of the external storage system using the namespace operator.
In the snapshot development step, the demonstration configures the snapshot of the backup data for utilizing the backup data.
In the data analytics step, the demonstration shows an example of data analytics using  backup data.

\subsection{Backup configuration}
\begin{figure}
	\centering
	\includegraphics[width=\linewidth]{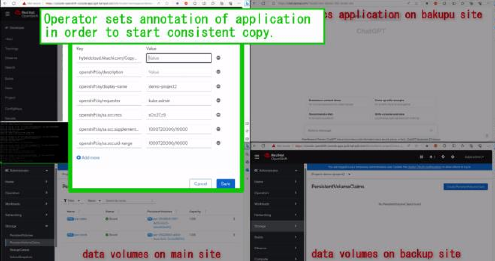}
	\caption{Operation in web console for configuring data backup of the business process.}
	\label{fig3}
\end{figure}

Fig.\ \ref{fig3} shows how the user initializes the data backup of a business process. 
The user tags the target namespace with the value ``ConsistentCopyToCloud,'' and then the namespace operator configures the ADC.

\begin{figure}
	\centering
	\includegraphics[width=\linewidth]{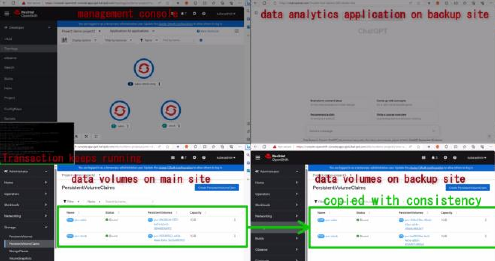}
	\caption{Change of persistent volume claims in the backup site.}
	\label{fig4}
\end{figure}

The lower left and right of the screen list the persistent volumes (PVs) in the main and backup sites, respectively.
Before tagging, the backup site had no PVs (Fig. \ref{fig3}).
Meanwhile, PVs appear in the backup site after tagging (Fig.\ \ref{fig4}).

The namespace operator automates the configuration of the ADC, including the consistency group setting.
This reveals the correspondence between PVs and storage volumes.
Therefore, it can complete the backup setting without knowledge of the external storage system.

\subsection{Snapshot development}
\begin{figure}
	\centering
	\includegraphics[width=\linewidth]{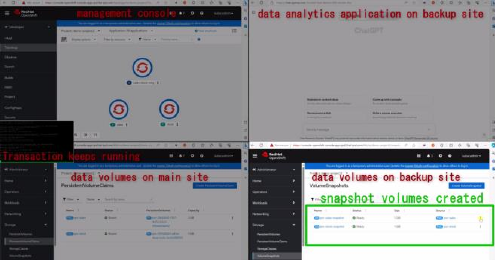}
	\caption{Snapshot volume development on the web console.}
	\label{fig5}
\end{figure}

Snapshot volumes are created on the backup site.
The operation is demonstrated in the lower right of the screen (Fig.\ \ref{fig5}).
Owing to the efforts in the storage plugins and container platform API, the development of snapshot can be completed on the web console.

The snapshot technology enables to access the volumes at the time of snapshot creation.
The demonstration system utilizes the snapshot volumes for the data analytics while data are being copied from the main to the backup sites.
Using storage technology, we can maintain consistency between the volumes at the main and backup sites.

\subsection{Data analytics}
\begin{figure}
	\centering
	\includegraphics[width=\linewidth]{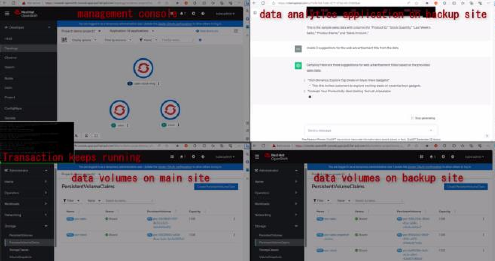}
	\caption{An example of data analytics using the snapshot volumes.}
	\label{fig6}
\end{figure}

Fig.\ \ref{fig6} shows an example of data analytics that uses snapshot volumes. 
Two databases are deployed in the backup site for reading snapshot volumes.
These databases provide backup data to the data analytics application.

\section{Related Work}
Synchronous data copy (SDC) is a prevailing method for data backup.
The proper usage and combination of SDC and ADC were investigated \cite{pu1991replica, goel2007data,natanzon2013dynamic,ciciani1990analysis}.
SDC has the advantage of no data loss, however the negative impact of SDC on business processing is inevitable due to the latency.

Phase locking is a fundamental technology for maintaining the consistency of distributed databases \cite{mohan1992aries, gardner2005phaselock}.
However, the ADC can impair the consistency of the backup data.
Thus, this paper employed the consistency group technologies, provided by the storage system, for maintaining the consistency in the system using the ADC.

\section{Conclusion}
Data backup in enterprise systems is required to minimize the impact on business processing, which can be categorized into two factors: system slowdown and downtime.
For eliminating both, the demonstration system employed the ADC and consistent group technologies.
We developed a namespace operator for automating the configuration of the ADC with a consistent group.
We integrated the storage and container technologies in the demonstration system, which can eliminate system slowdown and downtime.

\section*{Acknowledgment}
	We would like to thank the colleagues in Hitachi America ltd.\ for their supports in system and video developments.
	We would also like to thank Hitachi Vantara llc.\ for providing the system equipments and software. 

\bibliographystyle{IEEEtran}
\bibliography{mybibfile}

\end{document}